# Velocity Weakening in Anisotropic Friction on a Tilted Titania Nanorod Forest

Debottam Datta,[1] Enrico Gnecco,[2,*] J. P. Singh,[3,*] Nitya Nand Gosvami [4,*]

[1] School of Interdisciplinary Research, Indian Institute of Technology Delhi, Hauz Khas, New Delhi, 110016, India

[2] Marian Smoluchowski Institute of Physics, Jagiellonian University, 30348 Kraków, Poland

[3] Department of Physics, Indian Institute of Technology Delhi, Hauz Khas, New Delhi, 110016, India

[4] Department of Material Science and Engineering, Indian Institute of Technology Delhi, Hauz Khas, New Delhi, 110016, India

[*] Corresponding Author(s): enrico.gnecco@uj.edu.pl (Enrico Gnecco),

jpsingh@physics.iitd.ac.in (J. P. Singh),

ngosvami@mse.iitd.ac.in (Nitya Nand Gosvami)

## Abstract

In this study, we demonstrate velocity-dependent directional friction on a surface structured with tilted (~57°) titania nanorods using standard and colloidal probe force microscopy. Friction is measured at four different sliding speeds in two configurations, along and opposite to the tilt and perpendicular to the tilt direction, exhibiting anisotropic friction. Furthermore, friction decreases logarithmically with increasing sliding speed, which is attributed to the viscoelastic bending of the nanorods caused by stress-induced defect migration. The velocity weakening is more pronounced in the direction perpendicular to the tilt than along and opposite to it. The experimental findings are corroborated by creep measurements, which are well-reproduced by the Standard Linear Solid (SLS) model of viscoelasticity. Our results may be applied to the development of direction- and velocity-dependent sensors for microscale sliding motion as a robust alternative to structured interfaces based on polymeric materials.

Nano/microtextured surfaces control friction by tuning the real contact area,[1] interfacial adhesion,[2] energy dissipation,[3,4] stress distribution,[5] and tribochemical reactions[6] between sliding interfaces. Textures regulate speed-dependent friction[7,8] and reduce adhesion by disrupting continuous contact between tribo-pairs, which modifies the real contact area and capillary bridge formation.[9,10] These surfaces also generate ratchet movement,[11] exhibit anisotropic friction,[12,13] and delay the onset of wear by redistributing internal stress.[14] Owing to these outstanding features, nanotextured surfaces have found broad applications in developing low-stiction MEMS/NEMS,[15,16] superhydrophobic coatings,[17] microactuators,[18] robotic gripping and locomotion,[19,20] tactile sensors,[21] and biomedical interfaces.[22,23]

Although the tribological properties of nanotextured surfaces have been extensively studied, only a few investigations have addressed the issue of anisotropic and velocity-dependent friction on sub-micrometre scale textured surfaces. In this context, the reader is referred to the nanotextured surfaces of different materials developed using the glancing angle deposition (GLAD) technique, which exhibit anisotropic friction against various counter surfaces.[24-30] Furthermore, Baum et al.[31] fabricated a bio-inspired epoxy nanotextured surface and demonstrated frictional anisotropy when slid against a glass sphere. Pilkington et al.[32] observed velocity-weakening friction on various nanostructures, including zinc oxide nanograins, nanorods, and aluminium oxide nanodomes. Ando et al.[33] reported friction reduction with increasing sliding speed across different humidity conditions.

However, these studies were mainly focused on anisotropic or velocity-dependent friction, with only a limited number of studies examining both effects on the same surface. For instance, Cihan et al.[34,35] investigated the frictional properties of rippled PVS and patterned stainless steel (316L) using a PMMA sphere, finding frictional anisotropy on both surfaces, with velocity strengthening on PVS and velocity weakening on stainless steel. Extensions to other material pairs and a proper understanding of the underlying mechanisms are essential for translating this research into practical engineering applications.

In this study, a titania nanotextured surface was fabricated on a glass surface via the GLAD technique. FESEM images in Figure 1(a) and (b) present a cross-sectional and a top view of the nanotextured surface, respectively. The nanorods have an average length of 0.96 ± 0.03 μm and a diameter of 50 ± 15 nm, and they are tilted at α = 57° ± 4° relative to the substrate. Their structural properties were characterised using micro-Raman spectroscopy. Figure 1(c) portrays the micro-Raman spectra, revealing peaks at 149 cm$^{-1}$, 418.5 cm$^{-1}$, and 604.4 cm$^{-1}$, respectively, which correspond to the $B_{1g}$, $E_g$, and $A_{1g}$ vibrational modes of the rutile titania, with a significant redshift in the $E_g$ and $A_{1g}$ vibrational modes. An additional peak at 249.7 cm$^{-1}$ is attributed to multiple-phonon scattering due to the nanometer-scale structure of the nanorods.[36, 37] The redshift of the $E_g$ and $A_{1g}$ vibrational modes reveals characteristic changes in nanostructured titania. Swamy et al.[38] attributed the redshift in the $E_g$ and $A_{1g}$ vibrational modes to three-dimensional phonon confinement in reduced crystalline structures. Parker et al.[39] attributed a similar $E_g$ redshift in rutile titania to a reduced O/Ti stoichiometric ratio; accordingly, the redshift observed in Figure 1(c) indicates decreased O/Ti stoichiometry in the titania nanostructures. Nanostructure formation may introduce point defects, such as oxygen vacancies,[40, 41] which can be a primary reason for the decrease in the O/Ti stoichiometric ratio.

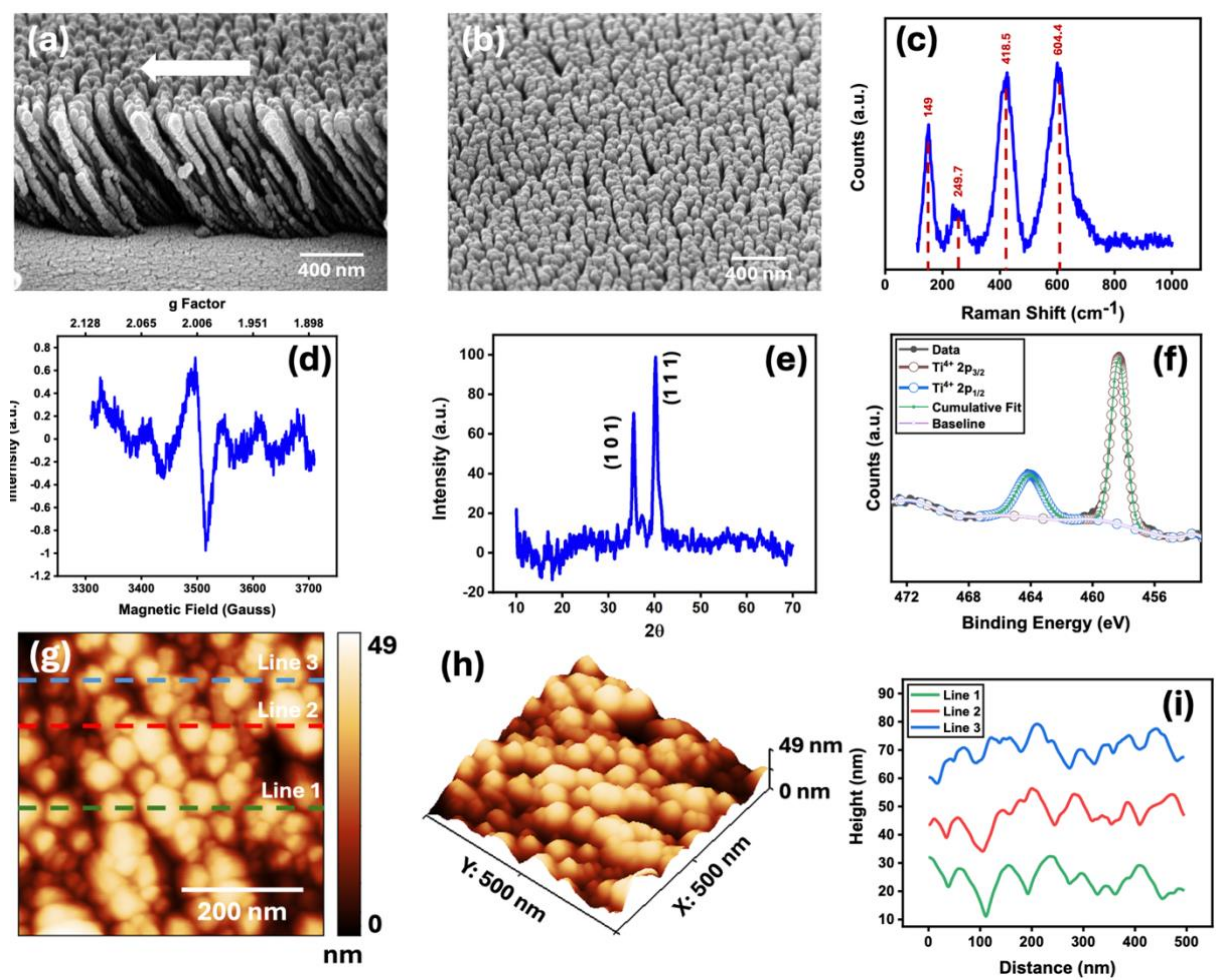

**Figure 1:** FESEM images presenting (a) a transverse section, and (b) a top view of the titania nanotextured surface (with the white arrow in figure a showing the direction of tilt); (c) micro-Raman; (d) EPR; (e) XRD and (f) XPS spectra of the surface; (g) 2D and (h) 3D view of an AFM topography of the surface; (i) Cross-sections corresponding to the dashed lines 1, 2 & 3 in g.

Electron paramagnetic resonance (EPR) spectroscopy was performed on X band to confirm the formation of point defects in the titania nanorods. The point defects, such as $Ti^{+3}$ centers, F-centers (single or double free electron trapped in oxygen vacancies), act as paramagnetic centers that can be detected using EPR spectroscopy.[42-44] In rutile titania, $Ti^{+3}$ centers exhibit a g-factor < 2.0, while it remains nearly ~2 for oxygen vacancies such as F-centers.[44] In Figure 1(d), the EPR spectrum of the nanotextures shows a g-factor of ~ 2.006, confirming the formation of the F-centers and the presence of oxygen vacancies in the titania nanorods. These vacancies are possibly responsible for the velocity-dependent frictional behaviour of titania nanorods, as discussed in detail later.

The titania nanorods were further characterised using X-ray diffraction (XRD, Cu Kα radiation, λ: 1.54 Å, Empyrean, Malvern PANalytical United Kingdom), and X-ray photoelectron spectroscopy (monochromatic X-ray source: AlKα 1486.8 eV, XPS, AXIS Supra, Kratos Analytical Ltd, United Kingdom). The XRD data in Figure 1(e) exhibit peaks at 35.5° and 40.6°, corresponding to the (1 0 1) and (1 1 1) planes of rutile titania (ICDD reference code 01-076-0325). In Figure 1(f), the XPS spectrum shows peaks $2p_{3/2}$ of $Ti^{4+}$ at 458.3 eV and $2p_{1/2}$ of $Ti^{4+}$ at 464.2 eV, confirming the chemical state and nanostructure of titania.[45] Representative surface topographies of the titania nanotextured surface were also captured using tapping-mode AFM. Figures 1 (g, h) show a 2D and the corresponding 3D topography of the textured surface. Figure 1(i) shows four-line profiles extracted from Figure 1(g). It reveals 7 to 9 peaks in a length of 500 nm, corresponding to an average peak-to-peak distance ($D_{p-p}$) of 63.1 ± 7 nm and an inter-rod spacing of ~15 ± 12 nm, consistent with FESEM observations in Figure 1(a,b).

Friction on the nanotextured surface was measured using lateral force microscopy (LFM) at ambient conditions. Before presenting the experimental results, we outline the protocol for measuring directional friction at various sliding velocities. Figures 2(a-d) schematically illustrate direction-dependent measurements using spherical alumina and a sharp DLC-coated AFM tip. Blue arrows denote the trace and retrace (back-and-forth) movement of the cantilever, while white arrows represent the tilt direction. Sliding was performed in two configurations: along and opposite to the tilt (AOT) and perpendicular to the tilt (PT).

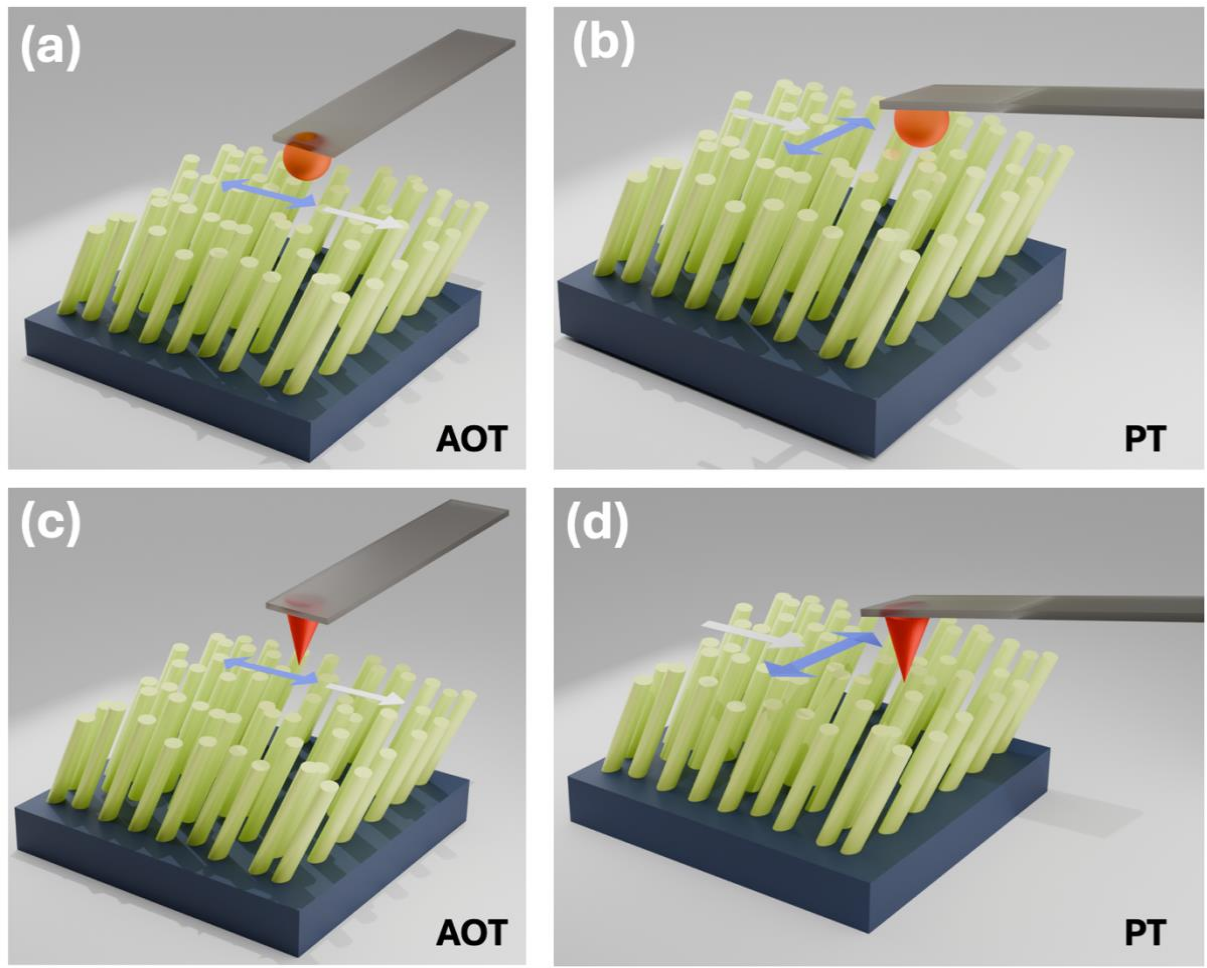

**Figure 2:** Schematic representation of directional friction measurements: The spherical alumina probe slides at (a) AOT and (b) PT configuration; and the DLC-coated AFM tip slides at (c) AOT and (b) PT configuration.

A first nanotribological characterization was performed at normal loads ($F_N$) of 600 nN and 800 nN using a ~30 μm spherical alumina probe in AOT and PT configurations at sliding speeds (V) of 5 μm/s, 7.14 μm/s, 16.66 μm/s, and 62.5 μm/s, respectively. Figures 3 (a, b) present the plot of the velocity dependence of the friction force, in a logarithmic scale, obtained in this way. Similar measurements were conducted using a sharp DLC-coated AFM tip at 50 nN and 100 nN, with the results shown in Figures 3 (c, d). In all cases, the friction force was found to be a linear function of log($V$):

$$F_f = a + b \times \log(V) \quad \text{--------------------------------(1)}$$

The values of $F_f$ intercept(*a*) and slope (*b*) are listed in Table 1 for all the measurements. We report remarkable findings: all slopes (*b*) are negative, indicating that the friction force decreases with increasing sliding speed in both AOT and PT configurations, regardless of probe type. Furthermore, the modulus of slope |*b*| is higher in the PT configuration compared to the AOT configuration, meaning that velocity weakening is more pronounced at the PT than at the AOT configuration. Moreover, the friction force is higher at AOT than in the PT configuration for almost all sliding velocities, regardless of whether a spherical alumina or a sharp AFM tip is used.

As discussed in previous studies,[46] the frictional anisotropy arises from the directional arrangement of titania nanorods. When the probe moves on the asymmetric pattern at the AOT position, it slides smoothly along the tilt direction but encounters substantial hindrance in the opposite direction. In contrast, the probe rubs symmetrically at the PT position, encountering less hindrance than at the AOT position. It is plausible that the nanorods are more compliant when rubbed at the AOT configuration than at the PT configuration, resulting in a larger contact area and higher friction at the AOT configuration.

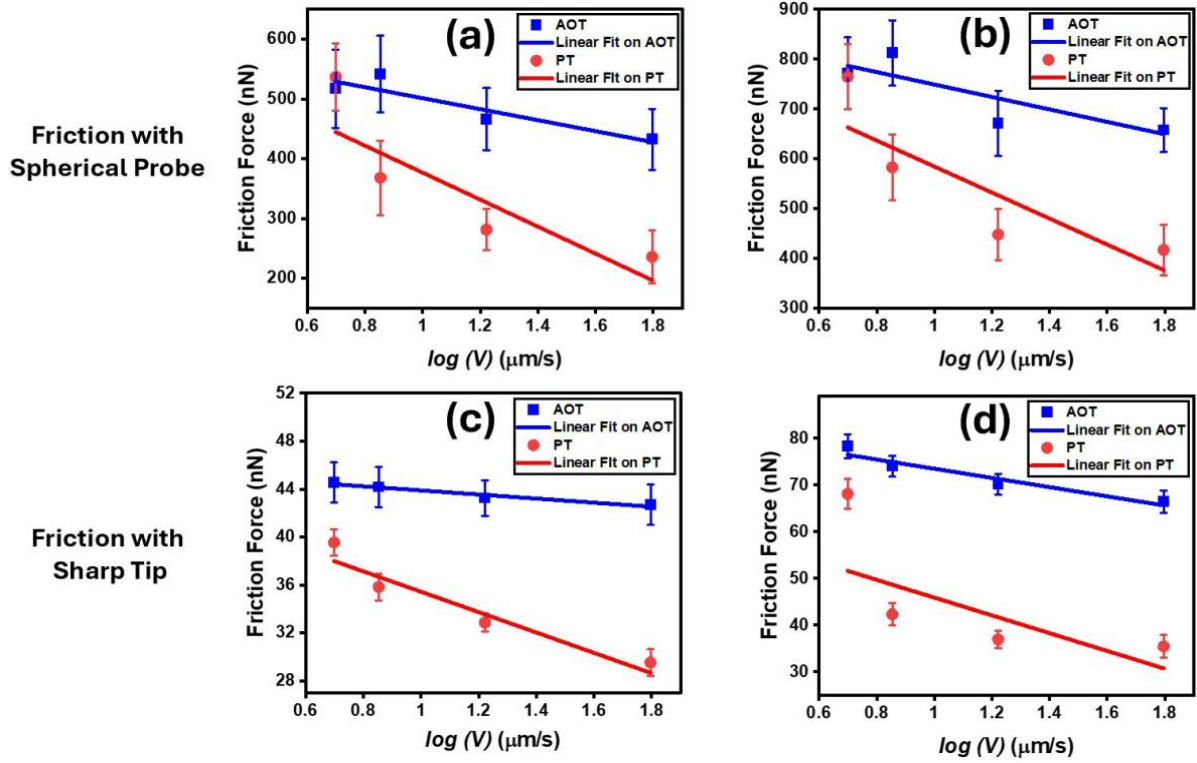

**Figure 3:** Friction force vs *log(V)* graphs obtained at the normal load of (a) 600 nN, (b) 800 nN, respectively, using spherical alumina probe; Friction force vs *log(V)* graphs obtained at the normal load of (c) 50 nN, (d) 100 nN, respectively, using sharp AFM tip.

**Table 1:** Values of $F_f$ intercept and slope of straight lines fitted to the $F_N$ vs *log (V)* plots

| Probe Type | Normal Load (nN) | AOT Configuration | | PT Configuration | |
|---|---|---|---|---|---|
| | | $F_f$ *Intercept (a)* (nN) | *Slope (negative) (b)* | $F_f$ *Intercept (a)* (nN) | *Slope (negative) (b)* |
| Spherical Alumina | 600 | 593.4 ±33.4 | 92 ± 26 | 603 ± 128.2 | 226 ± 98.9 |
| Spherical Alumina | 800 | 873.9 ± 67 | 125 ± 48.2 | 845.3 ± 148 | 261 ± 113 |
| Sharp Tip | 50 | 46 ± 0.37 | 1.7 ± 0.3 | 43.9 ± 2.2 | 8.5 ± 1.8 |
| Sharp Tip | 100 | 83.4 ± 2.5 | 9.8 ± 2.08 | 64.9 ± 18.1 | 19 ± 14 |

It remains to discuss why the titania nanotextured surface exhibits a logarithmic velocity weakening of friction. As mentioned earlier, Cihan et al.[34] observed a similar trend for a soft polymeric PMMA sphere sliding on a hard-textured steel surface. In this study, the effect was attributed to creep of PMMA (a polymer). At lower sliding velocities, the probe stays on the surface for longer, allowing creep to increase the real contact area and thus increase friction. Conversely, at higher velocities, reduced contact time limits creep, the contact area shrinks, and friction reduces for the PMMA-steel pair. In the present case, the observed decrease of friction with increasing sliding velocity on the "forest" of tilted titania nanorods can also be attributed to a time-dependent deformation of the nanorods. Each nanorod acts indeed as a compliant, anisotropic elastic element whose response is governed by bending rather than compression, owing to its high aspect ratio and preferential tilt direction. When the DLC-coated tip or the alumina microsphere slides across the surface at low velocity, the rods have sufficient time to bend, relax, and reattach to the slider in a quasi-static manner. As the sliding velocity increases, the characteristic excitation time imposed by the slider becomes shorter than the intrinsic retardation time of the nanorods. Under these conditions, the rods can no longer fully develop their bent equilibrium configurations. Instead, they are driven into a dynamically constrained state in which bending amplitudes are reduced, and unstable configurations are bypassed. This effectively suppresses energy losses, resulting in a lower friction force. In addition, the nanorod forest exhibits a rate-dependent effective stiffness. In section 2 of the supporting information, we discussed the velocity-dependent stick-slip behaviour observed on the titania nanotextured surface. In the AOT configuration, the occurrence of stick-slip is higher at lower velocities; however, it reduces significantly with increasing scan speed. In contrast to the AOT configuration, the number of stick-slip events is diminished at the PT configuration. At low velocities in asymmetric AOT sliding, the structure behaves as a mechanically soft layer, allowing many rods to deform sufficiently to participate in contact, thereby increasing the real contact area. At higher velocities, dynamic stiffening occurs: the rods respond more elastically and less dissipatively, reducing the number of rods that can sustain contact with the slider. The consequent reduction in real contact area further contributes to the observed velocity-weakening friction. The tilted geometry of the nanorods introduces a broken symmetry that amplifies this effect. Sliding induces asymmetric bending depending on the direction of motion relative to the tilt, and at higher velocities, the system preferentially follows lower-dissipation deformation pathways. Importantly, the similarity of the frictional behaviour observed with chemically and mechanically distinct sliders (DLC-coated tips and alumina microspheres) indicates that adhesion and interfacial shear play a secondary role. Instead, friction is governed primarily by bulk and surface energy dissipation within the nanorods, including internal friction and defect-mediated losses. The point defects, such as oxygen vacancies, formed on the GLAD-deposited, tilted titania nanotextured surface, are key contributors to this effect. When a bulk structure reduces to a nano- or microscale structure, it exhibits anelastic behaviour due to the motion of defects.[47] Chen et al.[48] and Cheng et al.[49] demonstrated that anelasticity can be observed in three-dimensional nanolattices due to stress-gradient-induced diffusion of point defects. In the rutile titania, anelasticity may arise due to stress-induced motion of point defects.[50] As discussed earlier, GLAD-deposited titania nanorods contain oxygen vacancies, confirmed by the micro-Raman and EPR spectra. So, the titania nanorod textured surface exhibits anelasticity or viscoelasticity upon sliding, which appears as the dominant factor for the emergence of velocity-weakening of friction.

To corroborate this interpretation, we have also performed nanoindentation measurements on the nanorod forest using a Berkovich indenter.[51] The indentation was performed using ramp loading to a maximum load of 0.1 mN at a constant loading rate. Once the indenter reached a specific depth under an applied load of 0.1 mN, the load was kept constant, and displacement was recorded for 120 seconds. Figure 4(a) and (b) present the plots of normal load ($P_N$) vs penetration depth ($d_p$) and normal load ($P_N$) vs indentation time ($t_p$), respectively. The green region in Figure 4(b) shows that the constant normal load ($P_N$) was maintained at 0.1 mN for 120 s. Figure 4(c) presents the plot of penetration depth ($d_p$) vs indentation time ($t_p$) for a single cycle of indentation (loading, holding, and unloading); the marked yellow region corresponds to the green region of Figure 4(b), exhibiting the displacement of the indenter at the constant load of 0.1 mN. The yellow-marked region is zoomed in, and Figure 4(d) presents the displacement of the indenter displacement at a constant load of 0.1 mN for 120 s. It shows an asymptotic increase in displacement over 120 seconds under constant load, confirming the viscoelastic behaviour of the titania nano-textured surface. The displacement was converted to creep compliance ($J$) using equation 2, which applies to the conical Berkovich indenter.[51]

$$J(t) = \frac{4 d_p^2(t)}{\Pi (1-\nu) P_N \tan \beta} \quad \text{------------------------------(2)}$$

Here, $\beta$ is the angle between the substrate plane and the cone generator, considered as 19.7°; $\nu$ is the Poisson's ratio of the Berkovich indenter, 0.04. It represents the time-dependent strain response of the nanorods under a constant applied stress.

Figure 4(e) presents a plot of $J$ vs indentation time ($t_p$), fitted with the function (Equation S1) based on the Standard Linear Solid (SLS) model,[51] where $\tau$ is the retardation time of nanorods. The indentation was performed at eight different places on the nanotextured surface, with the results in Table S2. The retardation time ($\tau$) is estimated, on average, as 41.3 ± 18 s. Formally, this is the same textbook behaviour reported for the viscoelastic deformation of polymeric materials.[52]

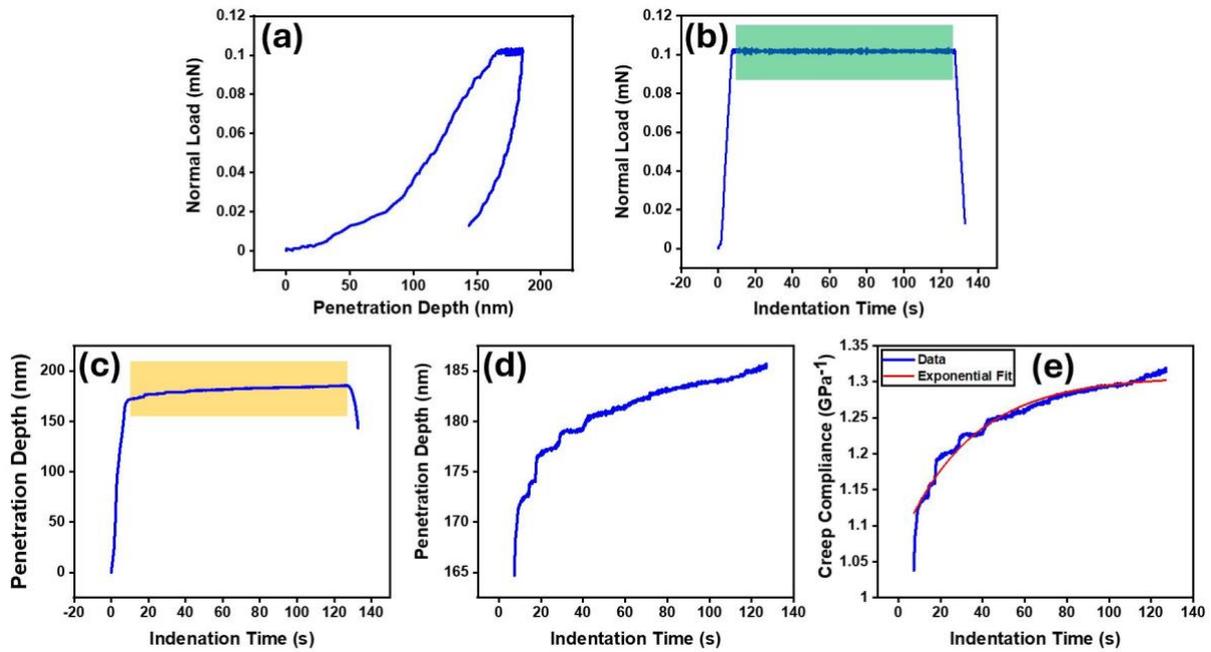

**Figure 4:** Nano-indentation measurement performed at the normal load of 0.1 mN to reveal the anelastic behaviour of the nanotextured surface; (a) Normal load ($P_N$) vs penetration depth ($d_p$) graph; (b) Normal load ($P_N$) vs indentation time ($t_p$) graph, where the green-marked region represents constant load of 0.1 mN for 120 s; (c) Penetration depth ($d_p$) vs indentation time ($t_p$) graph, where the yellow-marked region represents the penetration at the constant load of 0.1 mN for 120 s; (d) The zoomed-in graph of the yellow region marked in Figure c; (e) The graph of Figure d after exponential fit with creep compliance function.

Now, the question remains why the rate of velocity weakening is higher at the PT compared to the AOT orientation. This is probably due to the different bending dynamics of nanorods in the two configurations. A comprehensive understanding of this mechanism requires further investigation, and the underlying mechanisms remain open areas of research.

**Conclusion:**

In this work, we developed a titania nanorod textured surface on a glass substrate using the GLAD technique. Nanotribological tests were performed on the surface in two specific configurations, using both spherical alumina and a DLC-coated sharp AFM tip at four different combinations of sliding speed and normal load. The experimental data revealed anisotropic friction at almost all sliding velocities, due to the tilted arrangement of the nanorods. The nanotextured surface exhibited velocity weakening, meaning that friction decreased with increasing sliding velocity, with both colloidal and sharp AFM probes. The phenomenon occurred due to anelastic deformation of titania nanorods, driven by stress-induced movement of oxygen vacancies. The velocity-weakening was more pronounced in the PT configuration than in the AOT configuration. All in all, these results promote the proposed system as a viable alternative to structured polymeric surfaces for detecting microscale sliding motion in different directions.

**Supporting Information**

EDAX mapping of the titania nanotextured surface; Rate dependent stick-slip on titania nanotextured surface; Viscoelasticity in the nanotextured surface, Experimental Section: Preparation of textured surface, Measurement of nanoscale friction, Determination of viscoelastic behaviour using nanoindentation.

**Conflict of Interest**

The authors declare no competing financial interest.

**Acknowledgements:**

This research is sponsored by the Nano Mission Grant (DST/NM/NS/2019/289(G)) from the Department of Science and Technology (DST), India. The authors acknowledge the support of the Central Research Facility (CRF) and the Nanoscale Research Facility (NRF) of IIT Delhi. DD would like to express sincere gratitude to the Ministry of Education (MoE), India, for providing a full-time institute fellowship. E.G. acknowledges the support of the Strategic Program Excellence Initiative at the Jagiellonian University SciMat (Grant No. U1U/P05/NO/01.05).

**References:**

(1) Burton, Z.; Bhushan, B. Hydrophobicity, adhesion, and friction properties of nanopatterned polymers and scale dependence for micro- and nanoelectromechanical systems. *Nano Letters* **2005**, *5* (8), 1607-1613. DOI: 10.1021/nl050861b.

(2) Zou, M.; Cai, L.; Wang, H.; Yang, D.; Wyrobek, T. Adhesion and friction studies of a selectively micro/nano-textured surface produced by UV assisted crystallization of amorphous silicon. *Tribology Letters* **2005**, *20* (1), 43-52. DOI: 10.1007/s11249-005-7791-3.

(3) Zhang, H. S.; Komvopoulos, K. Scale-dependent nanomechanical behavior and anisotropic friction of nanotextured silicon surfaces. *Journal of Materials Research* **2009**, *24* (10), 3038-3043. DOI: 10.1557/Jmr.2009.0384.

(4) Sheng, Z. W.; Zhu, H.; He, Y.; Shao, B.; Sheng, Z.; Wang, S. Q. Tribological Effects of Surface Biomimetic Micro-Nano Textures on Metal Cutting Tools: A Review. *Biomimetics* **2025**, *10* (5). DOI: ARTN 28310.3390/biomimetics10050283.

(5) Park, J.; Kim, J.; Hong, J.; Lee, H.; Lee, Y.; Cho, S.; Kim, S. W.; Kim, J. J.; Kim, S. Y.; Ko, H. Tailoring force sensitivity and selectivity by microstructure engineering of multidirectional electronic skins. *Npg Asia Materials* **2018**, *10*, 163-176. DOI: 10.1038/s41427-018-0031-8.

(6) Xu, Y.; Li, Z. P.; Zhang, G. Q.; Wang, G.; Zeng, Z. X.; Wang, C. T.; Wang, C. C.; Zhao, S. C.; Zhang, Y. D.; Ren, T. H. Electrochemical corrosion and anisotropic tribological properties of bioinspired hierarchical morphologies on Ti-6Al-4V fabricated by laser texturing. *Tribology International* **2019**, *134*, 352-364. DOI: 10.1016/j.triboint.2019.01.040.

(7) Zhang, X. L.; Wang, X.; Kong, W.; Yi, G. W.; Jia, J. H. Tribological behavior of micro/nano-patterned surfaces in contact with AFM colloidal probe. *Applied Surface Science* **2011**, *258* (1), 113-119. DOI: 10.1016/j.apsusc.2011.08.015.

(8) Wakuda, M.; Yamauchi, Y.; Kanzaki, S.; Yasuda, Y. Effect of surface texturing on friction reduction between ceramic and steel materials under lubricated sliding contact. *Wear* **2003**, *254* (3-4), 356-363. DOI: 10.1016/S0043-1648(03)00004-8.

(9) Grewal, H. S.; Piao, S. X.; Cho, I. J.; Jhang, K. Y.; Yoon, E. S. Nanotribological and wetting performance of hierarchical patterns. *Soft Matter* **2016**, *12* (3), 859-866. DOI: 10.1039/c5sm01649e.

(10) Bhushan, B.; Jung, Y. C. Natural and biomimetic artificial surfaces for superhydrophobicity, self-cleaning, low adhesion, and drag reduction. *Progress in Materials Science* **2011**, *56* (1), 1-108. DOI: 10.1016/j.pmatsci.2010.04.003.

(11) Sekeroglu, K.; Gurkan, U. A.; Demirci, U.; Demirel, M. C. Transport of a soft cargo on a nanoscale ratchet. *Applied Physics Letters* **2011**, *99* (6). DOI: Artn 06370310.1063/1.3625430.

(12) Tramsen, H. T.; Gorb, S. N.; Zhang, H.; Manoonpong, P.; Dai, Z.; Heepe, L. Inversion of friction anisotropy in a bio-inspired asymmetrically structured surface. *J R Soc Interface* **2018**, *15* (138). DOI: 10.1098/rsif.2017.0629 From NLM Medline.

(13) Grewal, H. S.; Pendyala, P.; Shin, H.; Cho, I. J.; Yoon, E. S. Nanotribological behavior of bioinspired textured surfaces with directional characteristics. *Wear* **2017**, *384*, 151-158. DOI: 10.1016/j.wear.2017.01.033.

(14) Al-Azizi, A. A.; Eryilmaz, O.; Erdemir, A.; Kim, S. H. Nano-texture for a wear-resistant and near-frictionless diamond-like carbon. *Carbon* **2014**, *73*, 403-412. DOI: 10.1016/j.carbon.2014.03.003.

(15) Komvopoulos, K. Adhesion and friction forces in microelectromechanical systems: mechanisms, measurement, surface modification techniques, and adhesion theory. *Journal of Adhesion Science and Technology* **2003**, *17* (4), 477-517. DOI: Doi 10.1163/15685610360554384.

(16) Tayebi, N.; Polycarpou, A. A. Reducing the effects of adhesion and friction in microelectromechanical systems (MEMSs) through surface roughening: Comparison between theory and experiments - art. no. 073528. *Journal of Applied Physics* **2005**, *98* (7). DOI: Artn 07352810.1063/1.2058178.

(17) Tang, Y.; Yang, X. L.; Li, Y. M.; Lu, Y.; Zhu, D. Robust Micro-Nanostructured Superhydrophobic Surfaces for Long-Term Dropwise Condensation. *Nano Letters* **2021**, *21* (22), 9824-9833. DOI: 10.1021/acs.nanolett.1c01584.

(18) Li, S. P.; Yin, T. S.; Huang, Z. J. Soft Microstructure Arrays: A Multifunctional Microsystem for Mass Transport, Energy Transformation, and Surface Manipulation. *Acs Materials Letters* **2024**, *6* (5), 1686-1710. DOI: 10.1021/acsmaterialslett.4c00006.

(19) Daltorio, K. A.; Gorb, S.; Peressadko, A.; Horchler, A. D.; Ritzmann, R. E.; Quinn, R. D. A robot that climbs walls using micro-structured polymer feet. In *Climbing and Walking Robots: Proceedings of the 8th International Conference on Climbing and Walking Robots and the Support Technologies for Mobile Machines (CLAWAR 2005)*, 2006; Springer: pp 131-138.

(20) Li, S.; Tian, H. M.; Zhu, X. J.; Liu, M. X.; Li, X. M.; Shao, J. Y. Gecko Toe Pad-Inspired Robotic Gripper with Rapidly and Precisely Tunable Adhesion. *Research* **2025**, *8*. DOI: ARTN 068710.34133/research.0687.

(21) Chen, W.; Khamis, H.; Birznieks, I.; Lepora, N. F.; Redmond, S. J. Tactile Sensors for Friction Estimation and Incipient Slip Detection-Toward Dexterous Robotic Manipulation: A Review. *Ieee Sensors Journal* **2018**, *18* (22), 9049-9064. DOI: 10.1109/Jsen.2018.2868340.

(22) Ho, J. Y.; Rabbi, K. F.; Khodakarami, S.; Yan, X.; Li, L. N.; Wong, T. N.; Leong, K. C.; Miljkovic, N. Tunable and Robust Nanostructuring for Multifunctional Metal Additively Manufactured Interfaces. *Nano Letters* **2022**, *22* (7), 2650-2659. DOI: 10.1021/acs.nanolett.1c04463.

(23) Lee, M.-K.; Lee, H.; Kang, M.-H.; Hwang, C.; Kim, H.-E.; Oudega, M.; Jang, T.-S.; Jung, H.-D. Bioinspired nanotopography for combinatory osseointegration and antibacterial therapy. *ACS Applied Materials & Interfaces* **2024**, *16* (24), 30967-30979.

(24) Rodrigues, G. B.; Martin, N.; Amiot, F.; Colas, G. Friction anisotropy dependence on morphology of GLAD W films. *Tribology International* **2025**, *206*. DOI: ARTN 110556 10.1016/j.triboint.2025.110556.

(25) Mohanty, B.; Morton, B. D.; Alagoz, A. S.; Karabacak, T.; Zou, M. Frictional anisotropy of tilted molybdenum nanorods fabricated by glancing angle deposition. *Tribology International* **2014**, *80*, 216-221. DOI: 10.1016/j.triboint.2014.07.010.

(26) Mohanty, B.; Ivanoff, T. A.; Alagoz, A. S.; Karabacak, T.; Zou, M. Study of the anisotropic frictional and deformation behavior of surfaces textured with silver nanorods. *Tribology International* **2015**, *92*, 439-445. DOI: 10.1016/j.triboint.2015.07.027.

(27) So, E.; Demirel, M. C.; Wahl, K. J. Mechanical anisotropy of nanostructured parylene films during sliding contact. *Journal of Physics D-Applied Physics* **2010**, *43* (4). DOI: Artn 04540310.1088/0022-3727/43/4/045403.

(28) Hirakata, H.; Nishihira, T.; Yonezu, A.; Minoshima, K. Frictional Anisotropy of Oblique Nanocolumn Arrays Grown by Glancing Angle Deposition. *Tribology Letters* **2011**, *44* (2), 259-268. DOI: 10.1007/s11249-011-9844-0.

(29) Stempflé, P.; Besnard, A.; Martin, N.; Domatti, A.; Takadoum, J. Accurate control of friction with nanosculptured thin coatings: Application to gripping in microscale assembly. *Tribology International* **2013**, *59*, 67-78. DOI: 10.1016/j.triboint.2012.05.026.

(30) Pandey, M.; Verma, D.; Balakrishnan, V.; Gosvami, N. N.; Singh, J. P. Frictional anisotropy of Ag nanocolumnar surfaces. *Tribology International* **2021**, *153*. DOI: ARTN 10667410.1016/j.triboint.2020.106674.

(31) Baum, M. J.; Heepe, L.; Gorb, S. N. Friction behavior of a microstructured polymer surface inspired by snake skin. *Beilstein Journal of Nanotechnology* **2014**, *5*, 83-97. DOI: 10.3762/bjnano.5.8.

(32) Pilkington, G. A.; Thormann, E.; Claesson, P. M.; Fuge, G. M.; Fox, O. J. L.; Ashfold, M. N. R.; Leese, H.; Mattia, D.; Briscoe, W. H. Amontonian frictional behaviour of nanostructured surfaces. *Physical Chemistry Chemical Physics* **2011**, *13* (20), 9318-9326. DOI: 10.1039/c0cp02657c.

(33) Ando, Y. The effect of relative humidity on friction and pull-off forces measured on submicron-size asperity arrays. *Wear* **2000**, *238* (1), 12-19. DOI: Doi 10.1016/S0043-1648(99)00334-8.

(34) Cihan, E.; Heier, J.; Lubig, K.; Gräf, S.; Müller, F. A.; Gnecco, E. Dynamics of Sliding Friction between Laser-Induced Periodic Surface Structures (LIPSS) on Stainless Steel and PMMA Microspheres. *Acs Applied Materials & Interfaces* **2023**, *15* (11), 14970-14978. DOI: 10.1021/acsami.3c00057.

(35) Cihan, E.; Khaksar, H.; Lubig, K.; Gräf, S.; Müller, F. A.; Gnecco, E. Friction anisotropy in the sliding motion of polymer microspheres on a compliant rippled surface. *Physical Review E* **2025**, *111* (3). DOI: ARTN 03540510.1103/PhysRevE.111.035405.

(36) Suriani, A. B.; Muqoyyanah; Mohamed, A.; Othman, M. H. D.; Mamat, M. H.; Hashim, N.; Ahmad, M. K.; Nayan, N.; Khalil, H. P. S. A. Reduced graphene oxide-multiwalled carbon nanotubes hybrid film with low Pt loading as counter electrode for improved photovoltaic performance of dye-sensitised solar cells. *Journal of Materials Science-Materials in Electronics* **2018**, *29* (13), 10723-10743. DOI: 10.1007/s10854-018-9139-4.

(37) Ahmad, M. K.; Mokhtar, S. M.; Soon, C. F.; Nafarizal, N.; Suriani, A. B.; Mohamed, A.; Mamat, M. H.; Malek, M. F.; Shimomura, M.; Murakami, K. Raman investigation of rutile-phased $TiO_2$ nanorods/nanoflowers with various reaction times using one step hydrothermal method. *Journal of Materials Science-Materials in Electronics* **2016**, *27* (8), 7920-7926. DOI: 10.1007/s10854-016-4783-z.

(38) Swamy, V.; Muddle, B. C.; Dai, Q. Size-dependent modifications of the Raman spectrum of rutile TiO. *Applied Physics Letters* **2006**, *89* (16). DOI: Artn 16311810.1063/1.2364123.

(39) Parker, J. C.; Siegel, R. W. Calibration of the Raman-Spectrum to the Oxygen Stoichiometry of Nanophase Tio2. *Applied Physics Letters* **1990**, *57* (9), 943-945. DOI: Doi 10.1063/1.104274.

(40) Muhammad, P.; Zada, A.; Rashid, J.; Hanif, S.; Gao, Y.; Li, C.; Li, Y.; Fan, K.; Wang, Y. Defect engineering in nanocatalysts: from design and synthesis to applications. *Advanced Functional Materials* **2024**, *34* (29), 2314686.

(41) Yuan, M.; Kermanian, M.; Agarwal, T.; Yang, Z.; Yousefiasl, S.; Cheng, Z. Y.; Ma, P. A.; Lin, J.; Maleki, A. Defect Engineering in Biomedical Sciences. *Advanced Materials* **2023**, *35* (38). DOI: ARTN 230417610.1002/adma.202304176.

(42) Livraghi, S.; Rolando, M.; Maurelli, S.; Chiesa, M.; Paganini, M. C.; Giamello, E. Nature of Reduced States in Titanium Dioxide as Monitored by Electron Paramagnetic Resonance. II: Rutile and Brookite Cases. *Journal of Physical Chemistry C* **2014**, *118* (38), 22141-22148. DOI: 10.1021/jp5070374.

(43) Kumar, C. P.; Gopal, N. O.; Wang, T. C.; Wong, M. S.; Ke, S. C. EPR investigation of TiO nanoparticles with temperature-dependent properties. *Journal of Physical Chemistry B* **2006**, *110* (11), 5223-5229. DOI: 10.1021/jp057053t.

(44) Misra, S. K.; Andronenko, S. I.; Tipikin, D.; Freed, J. H.; Somani, V.; Prakash, O. Study of paramagnetic defect centers in as-grown and annealed $TiO_2$ anatase and rutile nanoparticles by a variable-temperature X-band and high-frequency (236 GHz) EPR. *Journal of Magnetism and Magnetic Materials* **2016**, *401*, 495-505. DOI: 10.1016/j.jmmm.2015.10.072.

(45) Chastain, J.; King Jr, R. C. Handbook of X-ray photoelectron spectroscopy. *Perkin-Elmer Corporation* **1992**, *40* (221), 25.

(46) Datta, D.; Gnecco, E.; Gosvami, N. N.; Singh, J. P. Anisotropic Stick–Slip Frictional Surfaces via Titania Nanorod Patterning. *ACS Applied Materials & Interfaces* **2024**. DOI: 10.1021/acsami.4c06428.

(47) Baker, S. P.; Vinci, R. P.; Arias, T. Elastic and anelastic behavior of materials in small dimensions. *Mrs Bulletin* **2002**, *27* (1), 26-29. DOI: DOI 10.1557/mrs2002.16.

(48) Chen, I. T.; Poblete, F. R.; Bagal, A.; Zhu, Y.; Chang, C. H. Anelasticity in thin-shell nanolattices. *Proceedings of the National Academy of Sciences of the United States of America* **2022**, *119* (38). DOI: ARTN e220158911910.1073/pnas.2201589119.

(49) Cheng, G. M.; Miao, C. Y.; Qin, Q. Q.; Li, J.; Xu, F.; Haftbaradaran, H.; Dickey, E. C.; Gao, H. J.; Zhu, Y. Large anelasticity and associated energy dissipation in single-crystalline nanowires. *Nature Nanotechnology* **2015**, *10* (8), 687-691. DOI: 10.1038/Nnano.2015.135.

(50) Carnahan, R.; Brittain, J. Point-defect relaxation in rutile single crystals. *Journal of Applied Physics* **1963**, *34* (10), 3095-3104.

(51) Lu, H.; Wang, B.; Ma, J.; Huang, G.; Viswanathan, H. Measurement of creep compliance of solid polymers by nanoindentation. *Mechanics of Time-Dependent Materials* **2003**, *7* (3-4), 189-207. DOI: 10.1023/B:MTDM.0000007217.07156.9b.

(52) Rösler, J.; Harders, H.; Bäker, M. *Mechanical behaviour of engineering materials: metals, ceramics, polymers, and composites*; Springer Science & Business Media, 2007.

# Supporting Information

## Velocity Weakening in Anisotropic Friction on a Tilted Titania Nanorod Forest

*Debottam Datta,[1] Enrico Gnecco,[2,*] J. P. Singh,[3,*] Nitya Nand Gosvami [4,*]*

[1] School of Interdisciplinary Research, Indian Institute of Technology Delhi, Hauz Khas, New Delhi, 110016, India

[2] Marian Smoluchowski Institute of Physics, Jagiellonian University, 30348 Kraków, Poland

[3] Department of Physics, Indian Institute of Technology Delhi, Hauz Khas, New Delhi, 110016, India

[4] Department of Material Science and Engineering, Indian Institute of Technology Delhi, Hauz Khas, New Delhi, 110016, India

*Corresponding Author(s): enrico.gnecco@uj.edu.pl (Enrico Gnecco),

jpsingh@physics.iitd.ac.in (J. P. Singh),

ngosvami@mse.iitd.ac.in (Nitya Nand Gosvami)

## 1. Characterization of Titania Nanorods:

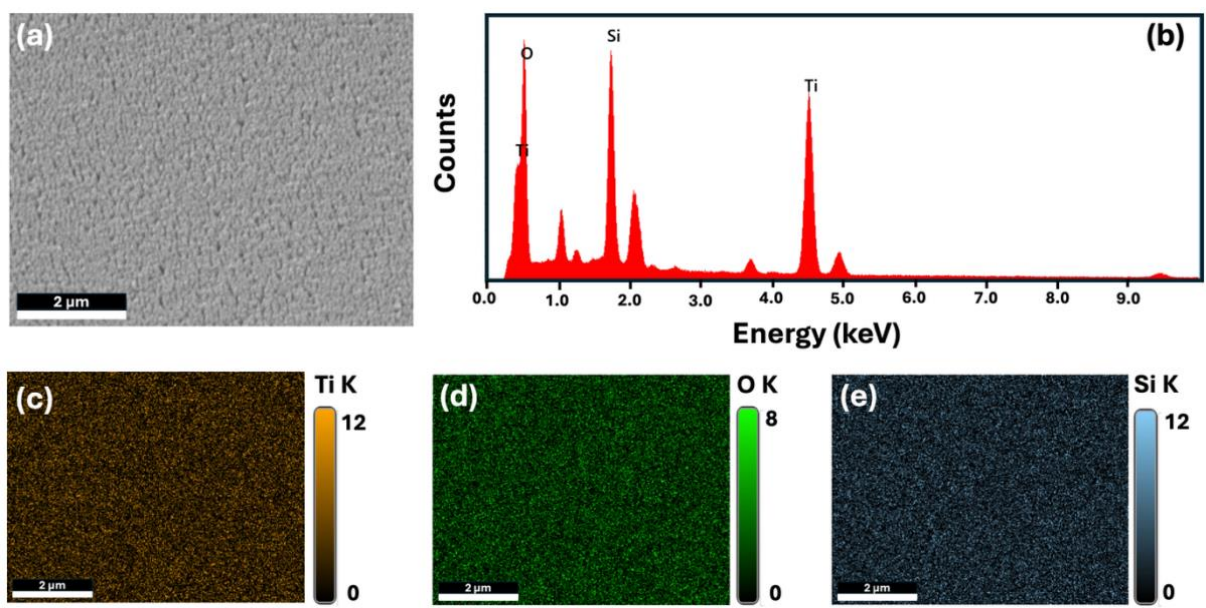

**Figure S1:** (a) FESEM image and (b) the Energy Dispersive Spectroscopy (EDS) analysis of the titania nanotextured surface; The elemental maps of (c) titanium, (d) oxygen, and (e) silicon.

**Table S1:** Elemental analysis of EDS conducted on the titania nanotextured surface

| Element | Weight % | Atomic % | Error % |
|---------|----------|----------|---------|
| Ti K    | 46.8     | 24.8     | 2.4     |
| O K     | 39.9     | 63.2     | 9.6     |
| Si K    | 13.3     | 12.0     | 3.7     |

The EDS analysis confirms the presence of titanium and oxygen on the nanotextured titania surface, as well as silicon in the glass substrate.

## 2. Rate-dependent Stick-Slip on Titania Nanotextured Surface:

The nanorod forest exhibited rate-dependent stick-slip at the AOT configuration during sliding with the spherical alumina probe. When sliding in the opposite direction to the tilt, stick-slip was prominent, and it reduced significantly when sliding along the tilt. However, the stick-slip was diminished at the PT configuration. Figures S2 and S3 demonstrate stick-slip behaviour at the AOT configuration under normal loads of 600 nN and 800 nN, respectively. The corrugated wavy features observed on the friction maps in Figures S2 and S3 correspond to the stick-slip (highlighted in blue) as shown in the trace (opposite to the tilt direction) curve in the friction loops. Interestingly, the number of stick-slip events is highest at the lowest sliding speed of 5 µm/s and gradually decreases with increasing velocity. Finally, the occurrence of stick-slip is significantly reduced at the velocity of 62.5 µm/s. Unlike the trace curve, the stick-slip event is suppressed at the retrace curve when sliding along the tilt direction.

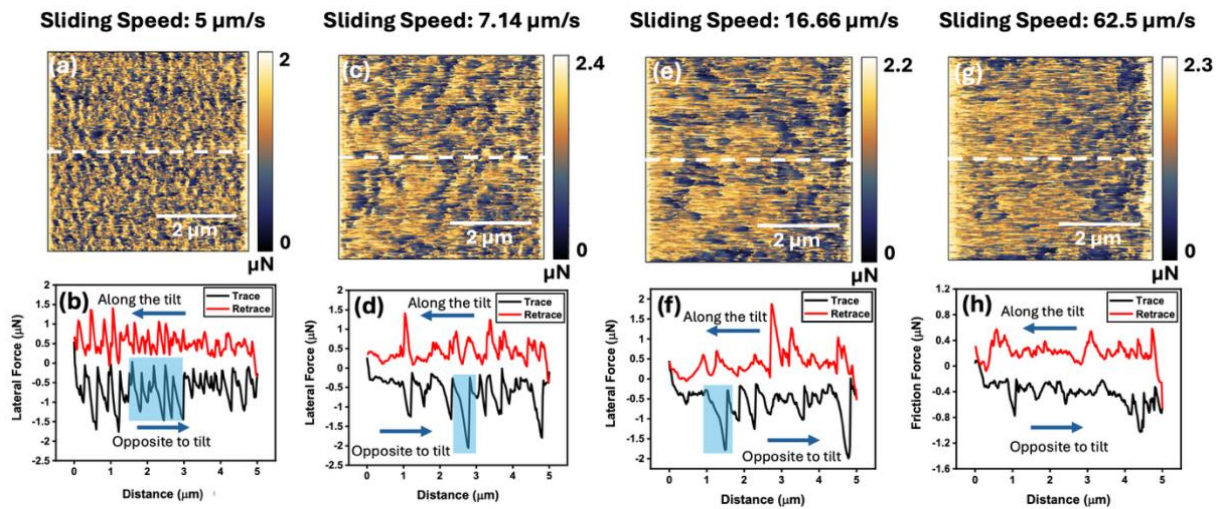

**Figure S2:** Evolution of friction maps as a function of increasing sliding velocity in the AOT configuration at the normal load of 600 nN using the spherical alumina probe. (a) Friction map and the (b) friction loop corresponding to the white dotted line marked in the map at the sliding speed of 5 µm/s; (c) Friction map and the (d) friction loop corresponding to the white dotted line marked in the map at the sliding speed of 7.14 µm/s; (e) Friction map and the (f) friction

loop corresponding to the white dotted line marked in the map at the sliding speed of 16.66 µm/s; (g) Friction map and the (h) friction loop corresponding to the white dotted line marked in the map at the sliding speed of 62.5 µm/s. The transparent blue box depicts the stick-slip feature. The trace and retrace curves in the friction loop represent sliding opposite and along the tilt directions, respectively.

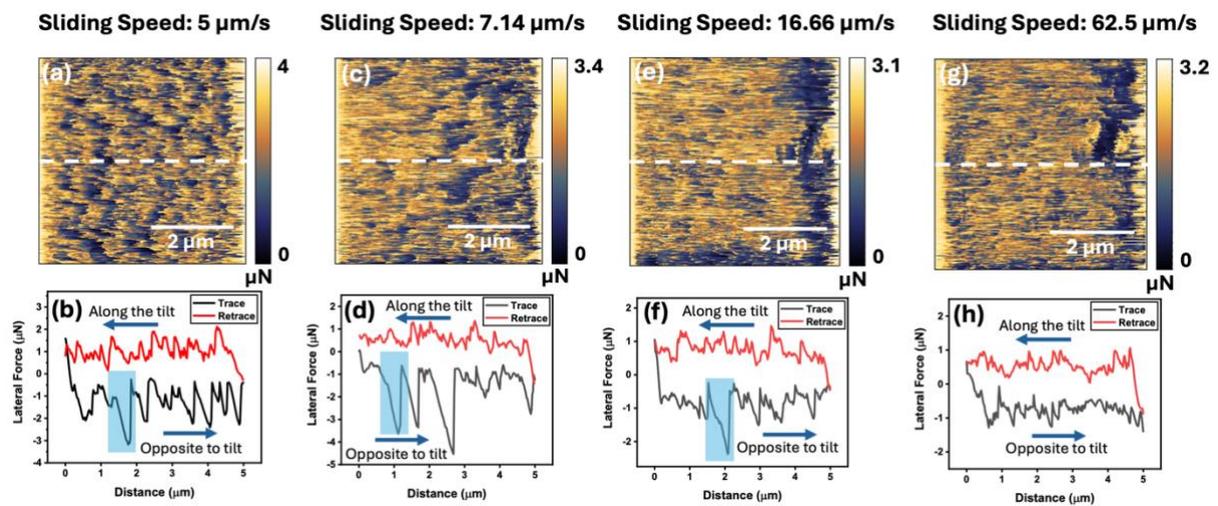

**Figure S3:** Evolution of friction maps as a function of increasing sliding velocity in the AOT configuration at the normal load of 800 nN using the spherical alumina probe. (a) Friction map and the (b) friction loop corresponding to the white dotted line marked in the map at the sliding speed of 5 µm/s; (c) Friction map and the (d) friction loop corresponding to the white dotted line marked in the map at the sliding speed of 7.14 µm/s; (e) Friction map and the (f) friction loop corresponding to the white dotted line marked in the map at the sliding speed of 16.66 µm/s; (g) Friction map and the (h) friction loop corresponding to the white dotted line marked in the map at the sliding speed of 62.5 µm/s. The transparent blue box depicts the stick-slip feature. The transparent blue box depicts the stick-slip feature. The trace and retrace curves in the friction loop represent sliding opposite and along the tilt directions, respectively.

In contrast to the AOT configuration, the stick-slip events are negligible in the PT configuration. Figures S4 and S5 show the friction maps and corresponding friction loops obtained at the PT configuration. The corrugated wavy features are also substantially reduced in the friction maps compared to those of the AOT configuration. This rate-dependent stick-slip behaviour significantly impacts the velocity-weakening behaviour of the nanorod forest. It is discussed in detail in the main manuscript.

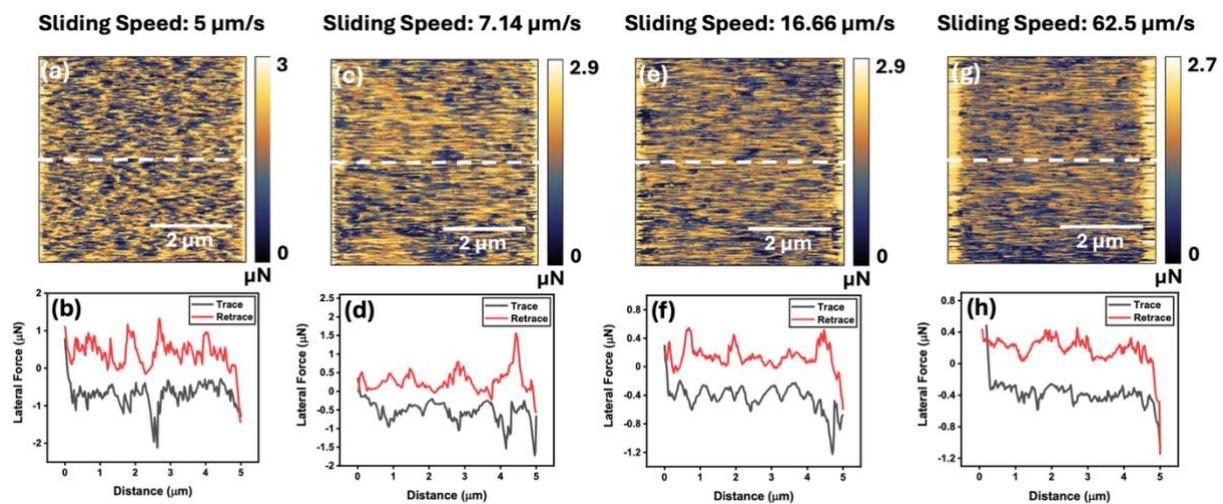

**Figure S4:** Evolution of friction maps as a function of increasing sliding velocity in the PT configuration at the normal load of 600 nN using the spherical alumina probe. (a) Friction map and the (b) friction loop corresponding to the white dotted line marked in the map at the sliding speed of 5 μm/s; (c) Friction map and the (d) friction loop corresponding to the white dotted line marked in the map at the sliding speed of 7.14 μm/s; (e) Friction map and the (f) friction loop corresponding to the white dotted line marked in the map at the sliding speed of 16.66 μm/s; (g) Friction map and the (h) friction loop corresponding to the white dotted line marked in the map at the sliding speed of 62.5 μm/s.

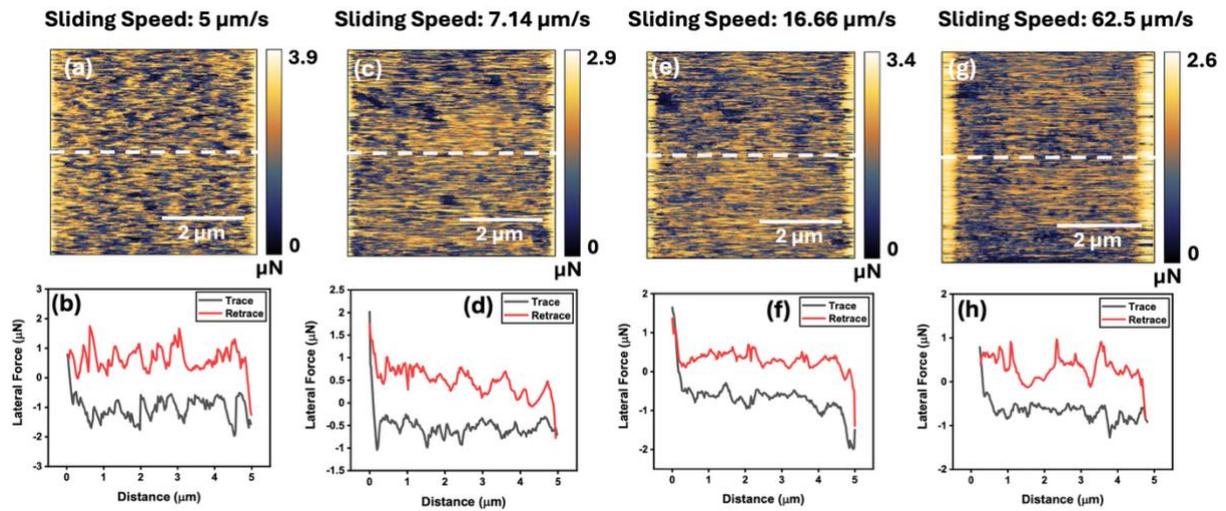

**Figure S5:** Evolution of friction maps as a function of increasing sliding velocity in the PT configuration at the normal load of 800 nN using the spherical alumina probe. (a) Friction map and the (b) friction loop corresponding to the white dotted line marked in the map at the sliding speed of 5 μm/s; (c) Friction map and the (d) friction loop corresponding to the white dotted line marked in the map at the sliding speed of 7.14 μm/s; (e) Friction map and the (f) friction loop corresponding to the white dotted line marked in the map at the sliding speed of 16.66 μm/s; (g) Friction map and the (h) friction loop corresponding to the white dotted line marked in the map at the sliding speed of 62.5 μm/s.

## 3. Viscoelasticity in Titania Nanotextured Surface:

Figure 4(e) presents the plot of creep compliance vs indentation time ($t_p$), which is fitted with the exponential creep function

$$J(t) = J_0 + J_1(1 - e^{\frac{-t_p}{\tau}})  \quad\quad\quad\text{------------------------(S1)}$$

based on the Standard Linear Solid (SLS) model.[1] Here, $J_0$, $J_1$ are the compliance numbers, and $\tau$ is the retardation time of nanorods. The indentation was performed at eight different places on the nanotextured surface, and the retardation time ($\tau$) is estimated as 41.2 ± 17.7 s. Table S2 shows the retardation time estimated at eight different locations.

**Table S2**: Retardation time calculated at eight different measurements

| Measurements | Retardation Time (seconds) |
|:---:|:---:|
| 1 | 17.27 |
| 2 | 54.82 |
| 3 | 56.32 |
| 4 | 61.23 |
| 5 | 33.4 |
| 6 | 14.5 |
| 7 | 46.9 |
| 8 | 45.93 |
| Average | 41.3 ± 18 |

The nanoindentation experiments confirm that the GLAD-deposited titania nanorod-textured surface exhibits viscoelastic behaviour, meaning the nanorods will undergo time-dependent deformation. When the AFM probe rubbed on the nanotextured surface, the nanorods would deflect accordingly.

## 4. Experimental Section:

### 4.1. Preparation of Titania Nanotextured Surface:

The titania nanotextured surface was prepared on a glass substrate using a physical vapour deposition (PVD)-based GLAD technique in an electron-beam (E-beam) evaporation system. Before undergoing the deposition, the glass substrates were first cleansed with labolene soap solution, then subjected to sequential ultrasonication in acetone, isopropyl alcohol (IPA), and DI water. The cleaned glass substrates were subsequently dried with a nitrogen jet.

The titanium pallets were loaded inside the graphite crucible and subsequently placed on the E-beam evaporation gun. To promote the growth of tilted titanium nanorods, the glass substrates were mounted on the sample holder at an angle of 81° relative to the vapour flux emanating from the graphite crucible, which was positioned on the E-beam gun. After reaching a high vacuum of $5 \times 10^{-6}$ Torr, E-beam evaporation was initiated and continued for 30 minutes, resulting in the tilted columnar growth of titanium nanorods. Due to the high reactivity of titanium in air, the metallic nanorods underwent oxidation and transformed to titanium dioxide ($TiO_2$).[2] The fabricated titania nanorods exhibited an average length of 0.96 ± 0.03 μm, diameters in the range of 50 ± 15 nm, and were inclined at an angle of ($\theta$) of 57° ± 4°. The FESEM images in Figures 1(a) and (b) portray the transverse section and top view of the titania nanotextured surface. The structural characterisation was conducted using micro-Raman, XRD, and XPS, confirming the formation of the rutile phase of titania.

### 4.2. Measurement of Friction Force on Titania Nanotextured Surface:

The friction force on the surface of the titania nanotextured surface was measured using the lateral force microscopy mode of a commercial AFM (Flex AFM, Nanosurf, Switzerland) in two different steps. First, the frictional measurement was performed after attaching a colloidal alumina of approximately ~30 μm diameter to a commercial AFM cantilever (SCM PIT) of stiffness 2 N/m. The sliding was carried out over a 5×5 μm² area in AOT and PT configurations, respectively, at scan speeds ranging from 5 μm/s to 62.5 μm/s, in increments of 7.14 μm/s and 16.66 μm/s, respectively. The experiment was performed at two normal loads: 600 nN and 800 nN, with the load remaining constant across all sliding speeds. Second, the same experiment was conducted using another DLC-coated sharp AFM tip (Multi 75DLC, Budget Sensors, Bulgaria) with a stiffness of 1.8 N/m. Similar to the colloidal alumina probe, the sliding measurement was conducted over a 5 × 5 μm² area in AOT and PT configurations at progressively increasing scan speeds of 5 μm/s, 7.14 μm/s, 16.66 μm/s, and 62.5 μm/s. The test was carried out in a batch process, discretising the experimental region into a 256 × 256-pixel grid. In each iteration, the AFM cantilever was slid back and forth, while the lateral deflection of the cantilever was simultaneously recorded for the trace (forward) and retrace (backwards) scans. The friction force was extracted by converting the lateral deflection signal and measuring the vertical half-width of the friction loops, which eliminated offsets from the zero horizontal axis caused by laser misalignment. The spring constants of both cantilevers were estimated using thermal tuning based on the Sadar and Green method.[3, 4] The acquired data were processed and analysed with a custom-made MATLAB code.

**4.3. Determining the Viscoelastic Behaviour of Titania Nanotextured Surface:**

The elastic modulus and anelastic behaviour of the titania nanotextured were determined using the nanoindentation technique. A Berkovich indenter was used for indentation, and the indenter penetrated the nanotextured surface until the normal load of 0.1 mN was applied. After achieving the normal load of 0.1 mN, the tip was held at this position to record deformation for 120 seconds, confirming the anelastic behaviour of the titania nanorods. The loading and unloading rate was 1 mN/min, and the approach and retract speed of the indenter was 1 μm/min. The indentation was performed at eight different places to estimate the elastic modulus and the creep compliance of the textured surface. Table S2 presents the estimated retardation time for the textured surface.

**References:**

(1) Lu, H.; Wang, B.; Ma, J.; Huang, G.; Viswanathan, H. Measurement of creep compliance of solid polymers by nanoindentation. *Mechanics of Time-Dependent Materials* **2003**, *7* (3-4), 189-207. DOI: 10.1023/B:MTDM.0000007217.07156.9b.

(2) Sarvadii, S. Y.; Gatin, A. K.; Kharitonov, V. A.; Dokhlikova, N. V.; Ozerin, S. A.; Grishin, M. V.; Shub, B. R. Oxidation of thin titanium films: Determination of the chemical composition of the oxide and the oxygen diffusion factor. *Crystals* **2020**, *10* (2), 117.

(3) Green, C. P.; Lioe, H.; Cleveland, J. P.; Proksch, R.; Mulvaney, P.; Sader, J. E. Normal and torsional spring constants of atomic force microscope cantilevers. *Review of Scientific Instruments* **2004**, *75* (6), 1988-1996. DOI: 10.1063/1.1753100 (accessed 9/2/2023).

(4) Sader, J. E.; Larson, I.; Mulvaney, P.; White, L. R. Method for the Calibration of Atomic-Force Microscope Cantilevers. *Review of Scientific Instruments* **1995**, *66* (7), 3789-3798. DOI: Doi 10.1063/1.1145439.